# Elemental Reactivity Maps for Materials Discovery

Yuki Inada[1*], Masaya Fujioka[2,3*], Haruhiko Morito[4], Tohru Sugahara[5], Hisanori Yamane[6], Yukari Katsura[7,8,9*]

[1] Department of Advanced Materials Science, The University of Tokyo, 5-1-5 Kashiwanoha, Kashiwa, Chiba, 277-8561, Japan

[2] Innovative Functional Materials Research Institute, National Institute of Advanced Industrial Science and Technology (AIST), 4-205 Sakurazaka, Moriyama-ku, Nagoya, Aichi, 463-8560, Japan.

[3] Research Institute for Electronic Science, Hokkaido University, Kita 20, Nishi 10, Kita-ku, Sapporo, Hokkaido, 001-0020, Japan

[4] Institute for Materials Research, Tohoku University, 2-1-1 Katahira, Aoba-ku, Sendai 980–8577, Miyagi, Japan

[5] Faculty of Materials Science and Engineering, Kyoto Institute of Technology, Matsugasaki, Sakyo-ku, Kyoto, 606-8585, Japan

[6] Institute of Multidisciplinary Research for Advanced Materials, Tohoku University, 2-1-1 Katahira, Aoba-ku, Sendai, Miyagi, 980-8577, Japan.

[7] Center for Basic Research on Materials, National Institute for Materials Science, 1-1 Namiki, Tsukuba, Ibaraki, 305-0044, Japan.

[8] Graduate School of Science and Technology, Tsukuba University, 1-1-1 Tennodai, Tsukuba, Ibaraki, 305-8573, Japan.

[9] RIKEN Center for Advanced Intelligence Project, RIKEN, 1-4-1 Nihonbashi, Chuo-ku, Tokyo, 103-0027, Japan.

*E-mail: ina980765263@gmail.com, m.fujioka@aist.go.jp, KATSURA.Yukari@nims.go.jp

## Abstract

When searching for novel inorganic materials, limiting the combination of constituent elements can greatly improve the search efficiency. In this study, we used machine learning to predict elemental combinations with high reactivity for materials discovery. The essential issue for such prediction is the uncertainty of whether the unreported combinations are non-reactive or not just investigated, though the reactive combinations can be easily collected as positive datasets from the materials databases. To construct the negative datasets, we developed a process to select reliable non-reactive combinations by evaluating the similarity between unreported and reactive combinations. The machine learning models were trained by both datasets and the prediction results were visualized by two-dimensional heatmaps: elemental reactivity maps to identify elemental combinations with high reactivity but no reported stable compounds. The maps predicted high reactivity (i.e., synthesizability) for the Co–Al–Ge ternary system, and two novel ternary compounds were synthesized: $Co_4Ge_{3.19}Al_{0.81}$ and $Co_2Al_{1.24}Ge_{1.76}$.

The development of new functional materials is an important aspect of applied materials science. Researchers in this field often spend a considerable amount of time working by trial and error to discover new materials. There is still no efficient way of searching for and synthesizing potentially useful novel functional materials. One of the main barriers in searching for new materials is the combinatorial explosion of the number of combinations of constituent elements. For example, considering the 80 elements that can be easily handled in laboratory experiments, ignoring the composition ratio, there are 3,160 possible binary elemental combinations (i.e., $_{80}C_2$). For ternary systems, there are 82,160 (i.e., $_{80}C_3$) combinations, and there are more than a million combinations for quaternary systems. Moreover, the number of combinations will be multiplied by the degrees of freedom of the composition ratio, such that the number of candidates becomes impractically large for trial-and-error approaches to synthesis. Some combinations of elements produce stable compounds in multiple composition ratios and with various crystal structures. Conversely, some combinations of elements do not react at all or do not produce stable compounds. Thus, from an experimental viewpoint, it is rare to successfully find a new material through a few experiments. Much effort is often required to search for potential combinations of constituent elements and identify appropriate compositions and reaction conditions. By reducing the large materials search space based on the reactivity of elemental combinations, the efficiency of discovering new materials will be greatly improved.

Recently, the field of materials informatics has emerged through the application of data science to materials datasets to improve the efficiency of materials research. Materials informatics is considered to be the fourth paradigm in materials science after experiment, theory, and simulation [1, 2].

The development of materials databases has enabled access to data about tens to hundreds of thousands of materials [3]. Large-scale crystal structure databases, such as the Inorganic Crystal Structure Database (ICSD) [4], The International Centre for Diffraction Data [5], and CRYSTMET [6], were established in the 1990s. Since the 2010s, various online databases based on *ab initio* calculations have become available, such as the Materials Project (MP) [7], AFLOW [8], and the Open Quantum Materials Database [9].

Attempts have been made to apply data science and machine learning techniques to such materials databases [10]. Machine learning can automatically learn the relationship between inputs and outputs from large amounts of data and make predictions. The first advantage of machine learning is that the predictions are fast. In many cases, machine learning models can make predictions for over thousands of datasets in a few seconds to minutes. The second advantage is that it is possible to output approximate prediction results based on limited information. For example, while crystal structure information is essential for *ab initio* calculations, machine learning enables predictions from more limited information, such as chemical compositions. Various properties have been predicted by machine learning using the composition as input, such as the formation energy [11–13], band gap [14], crystal structure [15, 16], superconducting transition temperature [17], and thermal conductivity [18]. Because of these features, machine learning is useful in searches for new materials, where predictions need to be made for a large number of unknown materials for which detailed data are unavailable.

Previous studies have attempted to use machine learning to propose compositions that might produce synthesizable compounds. Methods to predict the stability of compounds include the use of the predicted formation energy [13] and classifying whether the input compounds are stable or not [19]. When predicting the synthesizability of compounds from compositions as a classification problem, it is challenging to collect data on compositions that cannot be synthesized. It is difficult to determine in advance whether compounds that do not exist in databases are unstable compounds or are synthesizable but have been unreported. In other words, data that can be used in machine learning are positive data of stable compounds in the database, and unlabeled data that are not available in the database and are unknown whether they are positive or negative. The problem of machine learning with positive data and unlabeled data is known as positive unlabeled learning[20]. Seko *et al.* [19] proposed a recommendation system for ionic compounds by simple machine learning to discriminate compositional data existing in a database (positive data) and hypothetical compositional data (unlabeled data). Also, Vasylenko *et al.* [21] used a Variational Autoencoder to learn the characteristics of the elemental combinations with known compounds only from positive data and performed machine learning to suggest elemental combinations most likely to be synthesizable new compounds. Thus, material predictions have been developed and calculated using positive and unlabeled data or only using positive data. However, further accurate predictions should be achieved using positive and negative data. Some methods address the difficulty of obtaining true negative data by creating highly reliable negative data[22]. This study adopted utilizing reliable negative datasets by comparing the similarity of unlabeled and positive datasets; unlabeled data with low similarity to positive data is regarded as reliable negative data.

In addition, this study focused on the elemental combinations, not the chemical compositions. When predicting both the former and latter simultaneously, the search space becomes too large [19]. Therefore, ignoring the composition ratio and treating only the constituent elements is one effective option [21] since predicted elemental combinations can narrow the search space before considering compositional ratios. Also, such a prediction means the reactivity of each constituent element for synthesis.

The ternary elemental combinations were targeted, and neural networks (NNs) were trained to distinguish these positive and reliable negative data. The predicted reactivity for all of the ternary elemental combinations was visualized through heatmaps to provide the elemental reactivity at a glance. These elemental reactivity maps will allow researchers to greatly limit the search scope and easily identify elemental combinations where there is a high probability that novel materials can be obtained. Furthermore, information about the elemental reactivity can be used to discover elemental diffusion barrier materials. Thus, researchers in various fields will be able to determine their research plans more efficiently by referring to the elemental reactivity maps, in addition to their knowledge and experience. All of these elemental reactivity maps are openly available on the website. In addition, in this paper, examples of actual successful cases of new materials searching using these predicted results is presented.

# Method

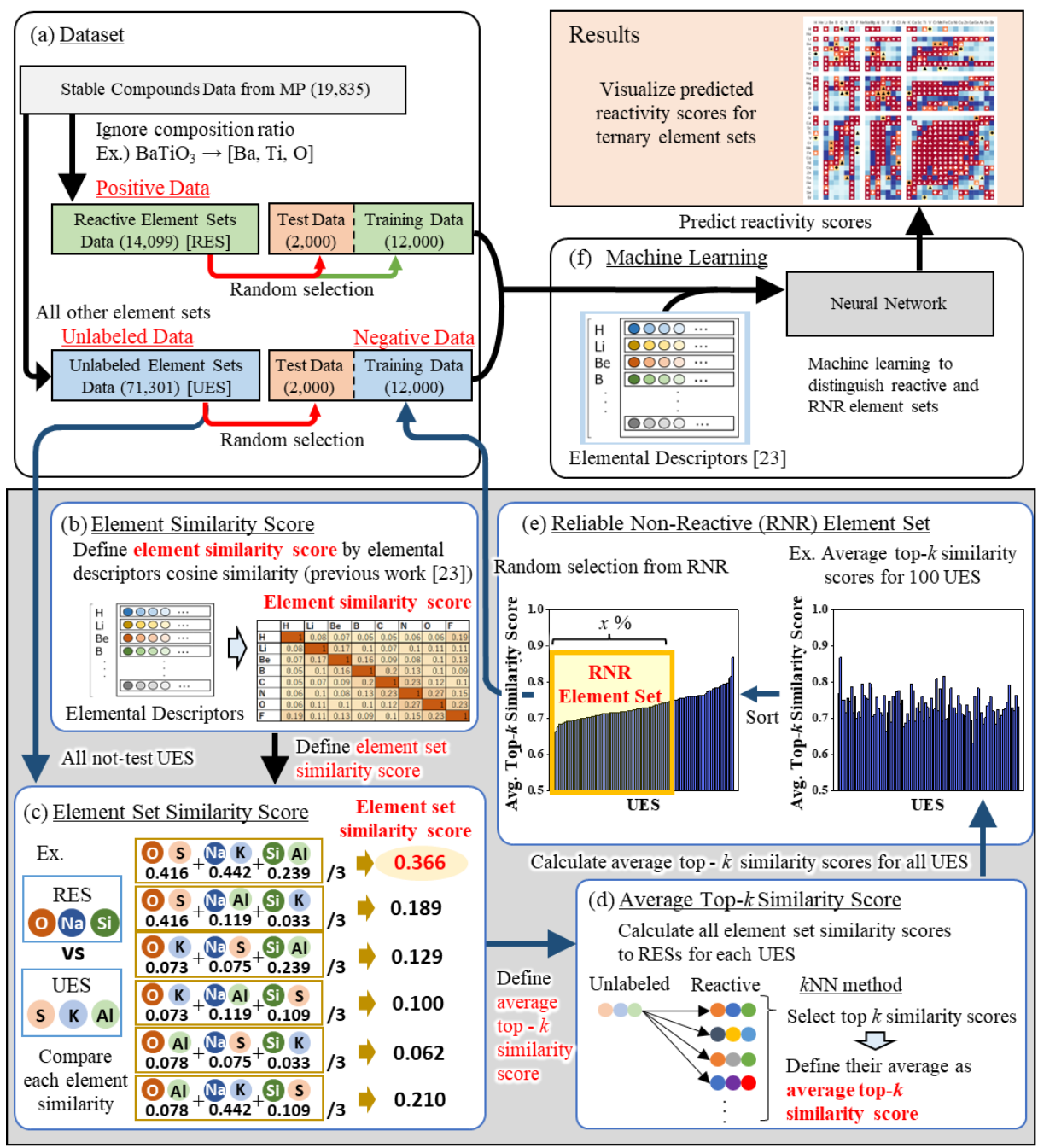

Figure 1. Overview of creation of elemental reactivity maps for ternary compounds. (a) Dataset obtained from the Materials Project (MP). The stable compound data from the MP are converted to reactive element set (RES) data. The remaining single, binary, and ternary element sets are unlabeled element sets (UESs). The test data of the RESs and UESs and the training data of the RESs are randomly selected. (b) Element similarity scores. The similarity of each element is defined by the cosine similarity of the element descriptors. (c) Element set similarity score between the RES and UES. (d) Average top-$k$ similarity score by the $k$NN method for each UES. (e) Reliable non-reactive element set (RNRES). The UESs with the bottom $x$% average top-$k$ similarity scores are regarded as the RNRESs. (f) Machine learning by a neural network using the elemental descriptors [23], RESs of the positive data, and RNRESs of the negative data. Prediction is performed for all single, binary, and ternary ESs, and it is visualized on heatmaps as elemental reactivity maps.

The flow of creation of the dataset and machine learning is shown in Figure 1(a)–(f). This study focuses on elemental combinations, and they are called element sets (ESs). An ES that produces some stable compounds is called a reactive element set (RES). RES data can be obtained from the chemical composition data of known stable compounds by ignoring the composition ratio (e.g., $BaTiO_3$ → [Ba, Ti, O]). From 19,835 computationally stable compounds from the MP (search performed April 25, 2018, energy_above_hull = 0), 14,099 ES data, including single, binary, and ternary ESs, were obtained. The other ESs are regarded as unlabeled element sets (UESs), which may or may not be reactive. When considering the 80 constituent elements that can be easily handled in the laboratory, excluding radioactive elements and noble gases, the sum of the number of single, binary, and ternary ESs is 85,400. Thus, the numbers of RESs and UESs were 14,099 and 71,301, respectively, as shown in Figure 1(a).

By random sampling from the 14,099 RESs and 71,301 UESs, two test datasets (Test-Dataset [RES] and Test-Dataset [UES], respectively) comprising 2,000 ESs were created. From the remaining RESs, 12,000 RESs were selected as positive training data (Training-Dataset [RES]). The remaining UESs were scored by comparing their similarity to the RESs through the processes shown in Figure 1(b)–(e). The UESs that were not similar to the RESs were regarded as reliable non-reactive element sets (RNRESs). Finally, 12,000 RNRESs were randomly selected as negative training data from all of the RNRESs.

To evaluate the similarity of the elements, the element similarity score is defined as the cosine similarity of the elemental descriptors [23], as described in the Supplementary Information (Figure S1). This similarity score between elements A and X is represented as CosSim(A, X). As a result, the score of each element's similarity is represented by a value between 0 and 1. The details are described in ref. 22.

The element set similarity score is defined as follows. Hereafter, the ES composed of elements A, B, ... is represented by ES[A, B, ...]. The similarity score between ES[A] and ES[X] is defined as CosSim(A, X). The similarity score between the binary system ES[A, B] and ES[X, Y] is set as max[CosSim(A, X) + CosSim(B, Y), CosSim(B, X) + CosSim(A, Y)]/2. In the case of the ternary systems ES[A, B, C] and ES[X, Y, Z], a total of six different X–Y–Z permutations ($X_i$–$Y_i$–$Z_i$ ($i$ = 1–6)) are created, and the maximum value of [CosSim(A, $X_i$) + CosSim(B, $Y_i$) + CosSim(C, $Z_i$)]/3 ($i$ = 1–6) is used as the element set similarity score. An example of the similarity between ES[O, Na, Si] and ES[S, K, Al] is shown in Figure 1(c).

$k$-nearest neighbor ($k$NN) scoring is a useful way of selecting reliable negative data from unlabeled data [22]. This approach was used to calculate the ES similarity scores for all of the RESs not in Test-Dataset [RES] for each UES, as shown in Figure 1(d). From these results, the top-$k$ similarity scores for each ES were averaged, which is defined as the average top-$k$ similarity score. In this study, $k$ was set to 4 through preliminary investigations, as described in the Supplementary Information (Figure S2 and Table S1).

The average top-$k$ similarity scores for all of the UESs except for Test-Dataset [UES] were sorted according to their score. The ESs with the average top-$k$ similarity scores in the bottom $x$% are regarded as the RNRESs. This operation is shown in Figure 1(e) for 100 UESs for ease of understanding. From the RNRESs, 12,000 ESs were randomly selected as negative data, and they are defined as Training-Dataset [RNRES ($x$%)]. Thus, the negative data, Training-Dataset [RNRES (*x*%)], and positive data, Training-Dataset [RES], were prepared for machine learning. In this study, we investigated four different models using *x* = 25, 50, 75, and 100.

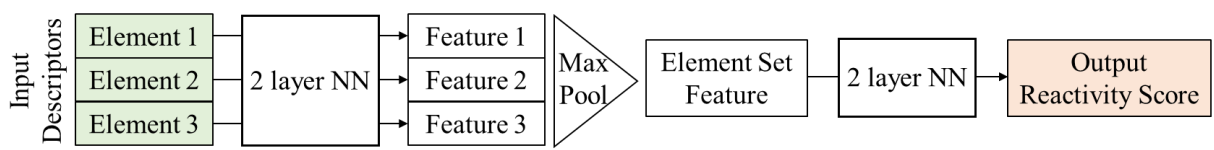

Figure 2. Structure of the neural networks used for machine learning and prediction.

The original structure of the NN used for machine learning is shown in Figure 2. First, the descriptors [23] (described in the Supplementary Information) of the three elements that constitute the ES were input and independently converted by a two-layer NN. When the number of elements in the ES was less than three, one of the same descriptors as the other input element was input as a dummy descriptor, and then the number of input elemental descriptors was set to three. The three vectors obtained from this conversion were max-pooled to return the same result regardless of the order of the input elements and the dummy elemental descriptors. This vector obtained by max-pooling was converted to the output value by a two-layer NN. The number of nodes of each hidden layer was set to 128. The activation functions were ReLU for the hidden layer and Sigmoid for the output layer. Cross-entropy was used as the loss function. AdamOptimizer [24] was used for training with a learning rate of 0.0001. The batch size was set to 100, and training was performed at 100 epochs.

The NN models were trained to output 1 for Training-Dataset [RES] and 0 for Training-Dataset [RNRES (*x*%)]. The input ESs are predicted to be more reactive for higher output values. For each training dataset, ten models were trained, and the average of their outputs was regarded as the reactivity score. These sets of models are represented as Model [*x*%] according to the *x* value of each Training-Dataset [RNRES (*x*%)].

## Results and Discussion

(i) Dataset for validation

The results predicted by the trained NN models were examined using two different datasets: Test-Dataset [RES, UES] obtained from the MP and the ICSD datasets not recorded in the MP. The RES datasets were prepared from computationally stable compounds in the MP as positive data, but not all of the reported compounds were covered. Namely, some materials, which are experimentally obtainable but are not recorded as stable compounds in the MP, were identified from the ICSD. These materials then provided 5,852 experimentally synthesizable ESs as ICSD datasets. Among these ESs, 2,578 ESs are categorized as high-quality data and the other 3,274 ESs are categorized as low-quality data in the ICSD, and they are labeled ICSD-HQ-Dataset and ICSD-LQ-Dataset in this study, respectively. Note that the ICSD data were searched on March 6, 2021.

Although it is not possible to experimentally synthesize the ternary compounds with high reactivity scores one by one to determine the prediction accuracy, the ICSD datasets can be regarded as 5,852 positive experimental results. Therefore, in this study, we used 5,852 ICSD datasets among the 71,301 UES datasets for prediction validation. Note that the UESs included the ESs from the ICSD datasets, so some may be used as training or test data.

(ii) Validation of the prediction results

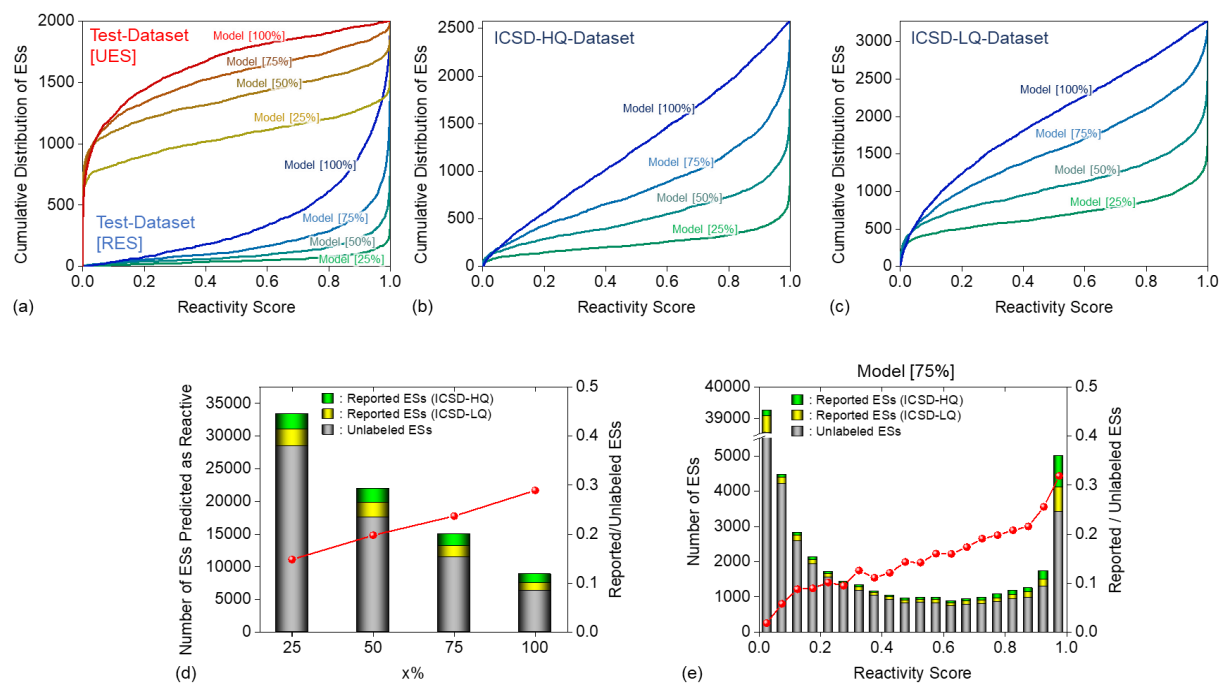

Figure 3. Validation of the predictions for Test-Dataset [reactive element set (RES)], Test-Dataset [unlabeled element set (UES)], and the ICSD datasets not recorded in the MP. All 71,301 UESs, including the training and test data, are classified into three data: ESs not reported in the MP but labeled as high-quality in the ICSD (ICSD-HQ-Dataset), ESs not reported in the MP but labeled as low-quality in the ICSD (ICSD-LQ-Dataset), and truly unlabeled ESs that are not reported in either the MP or ICSD. (a)–(c) Cumulative distributions of the number of ESs depending on the predicted reactivity scores for (a) Test-Dataset [RES] and Test-Dataset [UES], (b) ICSD-HQ-Dataset, and (c) ICSD-LQ-Dataset. As the reactivity score approaches 1, stable compounds can be obtained from the input element set (ES) with higher probability. The models trained using $x\%$ are called Model [$x\%$], where $x\%$ is defined as the threshold for selecting negative data. (d) Number of UESs predicted to be reactive by each Model [$x\%$] when reactivity score $\geq 0.5$ is regarded as reactive. The proportion of the correctly predicted ICSD datasets in the UESs predicted to be reactive is indicated by the red line. (e) Histogram of the number of UESs with respect to the reactivity score (0.05 increments) for all UESs (71,301). The proportion of the ICSD datasets in the UESs according to the reactivity score is indicated by the red line. In (d) and (e), ICSD-HQ-Dataset, ICSD-LQ-Dataset, and the other UESs are colored green, yellow, and gray, respectively.

The cumulative distributions of the prediction results for Test-Datasets [RES, UES], ICSD-HQ-Dataset,

and ICSD-LQ-Dataset are shown in Figure 3(a)–(c), respectively. When the cumulative distribution curve approaches the bottom right corner, it indicates an increasing number of ESs with high reactivity scores. Conversely, when the curve approaches the top left corner, it indicates a higher frequency of low reactivity scores. There were different trends for Test-Dataset [RES] and Test-Dataset [UES], showing the correct prediction that the RESs are much more reactive than the UESs (Figure 3(a)). In addition, the intermediate trends between Test-Dataset [RES] and Test-Dataset [UES] were confirmed from the results for ICSD-HQ-Dataset and ICSD-LQ-Dataset in Figure 3(b) and (c). Despite the ICSD datasets being classified as UESs, their cumulative distributions differed from that of Test-Dataset [UES]. Furthermore, for all of the datasets, lower $x$% resulted in the cumulative curves being closer to the bottom right corner, that is, the number of ESs with high reactivity scores increased for lower $x$%.

When ESs with reactivity score ≥ 0.5 were regarded as reactive (threshold = 0.5), Model [25%] correctly predicted that 98%, 91%, and 80% of Test-Dataset [RES], ICSD-HQ-Dataset, and ICSD-LQ-Dataset were reactive, respectively. For each dataset, this percentage decreased for higher $x$%: 94%, 71%, and 53% for Model [75%], and 88%, 52%, and 37% for Model [100%], respectively. A possible reason for this trend is that as $x$% increases, more ESs are erroneously regarded as negative data in the process in Figure 1(e). However, these results do not mean that lower $x$% improves the prediction accuracy of the true synthesizability. It can be interpreted that low $x$% makes more optimistic predictions than high $x$%; the number of ESs predicted to be reactive increased for lower $x$%, regardless of RES and UES. For example, in the case of Model [25%] using threshold = 0.5, the proportion of ESs predicted to be reactive of all of the UESs was 47%. This proportion decreased to 31%, 20%, and 12% when $x$% was increased to 50%, 75%, and 100%, respectively. Namely, the number of ESs predicted to be reactive differed by approximately four times between Model [25%] (33,403 ESs) and Model [100%] (8,853 ESs). Thus, higher $x$% tends to reduce the search space for new compounds. As a result, the proportion of ICSD datasets that were correctly predicted to be reactive of all of the UESs predicted to be reactive was higher for higher $x$%: 15%, 20%, 24%, and 29% for Model [25%], [50%], [75%], and [100%], respectively, as shown in Figure 3(d). Therefore, this trend is considered to be the actual probability of finding new materials, showing the importance of using high $x$%.

The histogram of the change in the number of ESs among the 71,301 UESs and the line graph of the ratio of the ICSD datasets in the UESs with respect to the reactivity score for Model [75%] are shown in Figure 3(e). The results for the other models, Model [25%], [50%], and [100%], are shown in Figure S3. The ICSD datasets/UESs ratio tended to increase with increasing reactivity scores, as highlighted in red. Thus, the prediction results demonstrate that the obtained reactivity scores reflect the actual synthesizability and higher reactivity scores could lead to higher probability of discovering new materials.

(iii) Discussion on the results of Test-Dataset [RES] and the ICSD datasets

As described above, the intermediate trends between Test-Dataset [RES] and Test-Dataset [UES] were confirmed from ICSD-HQ-Dataset and ICSD-LQ-Dataset in Figure 3(a)–(c). Even though both ICSD-HQ-Dataset and ICSD-LQ-Dataset are experimentally synthesizable, as the data quality decreases, the trend becomes closer to that of Test-Dataset [UES]. It is not easy to investigate the exact reason for these differences because the MP and ICSD contain tens of thousands of data, but a possible reason is given below.

Comparing Test-Dataset [RES] and the ICSD datasets, the former is composed of ESs with stable compounds in the MP, while the latter datasets are composed of ESs from experimentally obtained compounds excluding the computationally stable materials in the former. For instance, solid solutions, which are difficult to investigate by first-principles calculations, and metastable compounds, which were omitted from the positive data owing to their instability, are considered to be included in the ICSD datasets. Because only computationally stable compounds were regarded as reactive in machine learning, this may be the reason for the bias between Test-Dataset [RES] and the ICSD datasets, leading to their different behaviors.

In addition, comparing ICSD-HQ-Dataset and ICSD-LQ-Dataset, the latter with lower crystallinity should include more metastable or difficult-to-synthesize materials than the former. Therefore, it is considered that the latter is more affected by the above-mentioned tendency than the former. In other words, the predictive process developed in this study may be good at finding more stable compounds.

(iv) Selection of an appropriate $x$%

From the results of validation with Test-Dataset [RES], Test-Dataset [UES], and the ICSD datasets, the predicted reactivity scores were highly dependent on the value of $x$%. It is not easy to determine the optimal $x$% in advance because of the advantages and disadvantages depending on $x$%. Each researcher should select an appropriate $x$% according to the purpose of their experiments. Models with higher $x$% predict fewer ESs with high reactivity scores and help to decrease the search space for new compounds. In addition, higher $x$% can provide higher probability of finding the RESs than lower $x$%, as shown in Figure 3(d). However, some ESs that are reactive are regarded as not reactive in models with high $x$%, which may lead to overlooking potential undiscovered materials. Conversely, models with low $x$% are not good at decreasing the search space, but they are less likely to overlook RESs. From another viewpoint, low $x$% models may be useful to identify a combination of elements that is not likely to react and assist with finding barrier materials that prevent diffusion owing to their unreactive features.

(v) Visualization of the predicted reactive scores on elemental reactivity maps

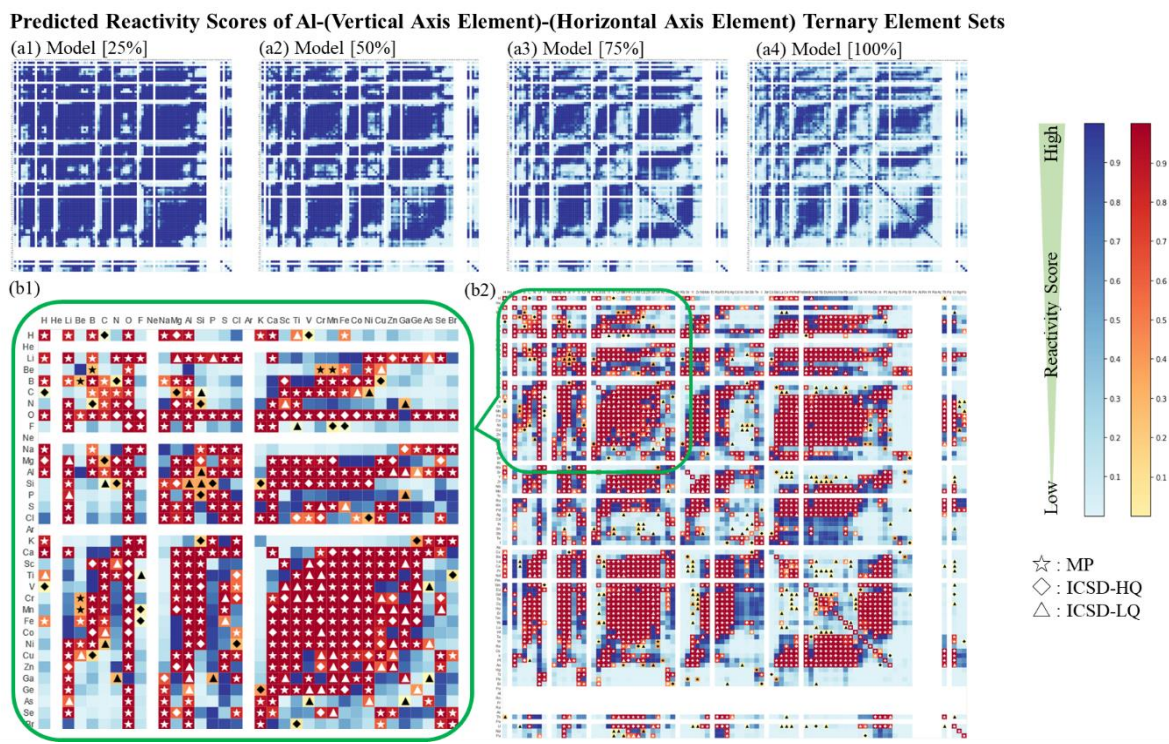

Figure 4. Examples of elemental reactivity maps visualizing the predicted results. The main target element of these maps is Al. The elements from H to Pu are arranged on the vertical and horizontal axes, and each square of the heatmaps represents the predicted reactivity score of element sets (ESs) comprising Al, the vertical axis element, and the horizontal axis element. (a1)–(a4) Elemental reactivity maps showing only the predicted reactivity scores. The reactivity score is higher for deeper colors. (b) Elemental reactivity map of (a3) along with the ESs for known compounds in the Materials Project (MP) and Inorganic Crystal Structure Database (ICSD). (b1) is an expanded view of (b2). The ESs of known compounds are highlighted in red. Known compounds are classified by the source of the data. Stars indicate the stable compounds in the MP. Diamonds and triangles indicate compounds not in the MP but in the ICSD, and they are labeled as high and low quality from the ICSD. These symbols are white if the predicted reactivity score is 0.5 or higher and black if it is less than 0.5. Note that only data from the MP are used as positive data for the reactive element set in this study, so data from the ICSD are treated as unlabeled element sets when machine learning, even though they are highlighted in red on the heatmaps.

The results predicted by the trained models were visualized by two-dimensional heatmaps (elemental reactivity maps). Each elemental reactivity map shows the predicted reactivity scores of ternary element sets including the target element. The maps with the target element Al for Models [25%–100%] are shown in Figure 4(a1)–(a4). A deeper blue color of the square corresponds to a higher reactivity score. The number of ESs with high reactivity scores obviously increased with decreasing $x$%. The maps for Model [75%] are shown in Figure 4(b) owing to the balance between the size of the search space and the certainty of prediction. The red squares in Figure 4(b) correspond to the reported data in the MP or ICSD. Therefore, using elemental reactivity maps, researchers in the inorganic field can easily determine ESs that have not been reported but are highly synthesizable (dark blue squares) to discover novel materials. All of the elemental reactivity maps for the 80 different target elements are provided at https://erm.starrydata.org/.

(vi) Experimental results using material search maps

In this section, we describe the discovery of new ternary compounds based on the elemental reactivity maps. In Figure 3(e), there are 5,006 ESs with reactivity scores greater than 0.95. From these 5,006 ESs, a series of ESs with chemical stability, easy availability, and non-toxicity was chosen: [$M$, Al, Ge] ($M$ = Ti, V, Cr, Mn, Fe, Co, Ni). Among the ESs of [(Ti, Mn, Ni), Al, Ge], some ternary compounds have been recorded in the MP, such as $Al_3GeTi$ [25], Mn–Al–Ge, quasicrystals [26], and $CoGe_2/CaF_2$-type Ni–Al–Ge ternary alloys [27], and these ESs are marked in red in Figure S4. The remaining ESs of [(V, Cr, Fe, Co), Al, Ge] correspond to UESs and show high reactivity scores, such as V = 0.9927, Cr = 0.9697, Fe = 0.9979, and Co = 0.9866, and they correspond to dark blue in the elemental reactivity map. Although the materials belonging to these ESs are not included in the MP, ternary compounds such as $V(Al/Ge)_2$ [28], CrAlGe [29], and $Al_{3-x}FeGe_{2+x}$ [30] are reported as experimentally synthesized materials in the ICSD. Therefore, [Co, Al, Ge] is the only combination that has not been reported and has a high reactivity score for [$M$, Al, Ge], as shown in Figure S4.

In this study, two new compounds in the Co–Al–Ge ternary system were successfully synthesized by simple solid-state reaction: B20-type structure $Co_4Ge_{3.19}Al_{0.81}$ and $Ni_2Al_3$-like structure $Co_2Al_{1.26}Ge_{1.74}$. The experimental details are described in Section 4 in the Supplementary Information. The results of crystal structure analysis for single crystals of these two compounds are given in Table S1. The crystal structures of $Co_4Ge_{3.19}Al_{0.81}$ and $Co_2Al_{1.26}Ge_{1.74}$ are shown in Figure 5. The details of the experimental procedure and crystal structure information are provided in the Supplementary Information.

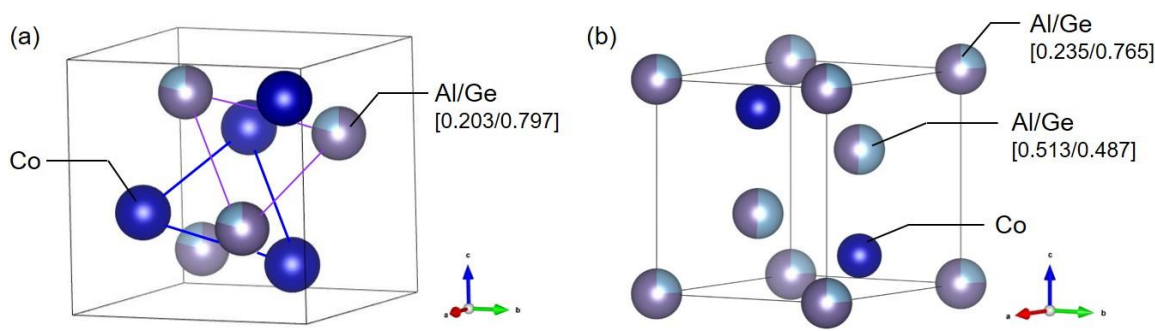

Figure 5. Crystal structures of (a) $Co_4Ge_{3.19}Al_{0.81}$ and (b) $Co_2Al_{1.24}Ge_{1.76}$. The site occupancies of Al and Ge are shown in brackets. The structures were drawn by the VESTA program [31].

## Conclusion

In this study, we developed elemental reactivity maps, focusing on ternary systems. These maps provide information about whether the three selected elements among 80 elements easily handled in the laboratory are reactive. This information can be used depending on the research purpose, such as synthesizing new ternary compounds or finding effective element combinations for diffusion barriers.

One of the difficulties in predicting the reactivity using machine learning is determining the negative data, because it is unknown whether unreported materials are not synthesizable or have just not been investigated. To address this issue, the similarity of the unlabeled data to the positive data was calculated through the process shown in Figure 1(b)–(e). The lowest $x\%$ similarity scores were considered to be reliable non-reactive data (negative data).

NN models were trained by using these training data. According to the $x\%$ value, the prediction results showed different trends, and the advantages and disadvantages of each $x\%$ value were carefully discussed. Lower $x\%$ models tended to make more optimistic predictions. This is not good for decreasing the material search space, but it is less likely to overlook reactive elemental combinations. Conversely, higher $x\%$ models provide fewer candidates with high reactivity scores. Therefore, it helps to decrease the material search space, but it may lead to potential undiscovered materials being overlooked. Thus, an appropriate $x\%$ value should be selected considering the characteristics of $x\%$.

The effectiveness of the developed elemental reactivity maps for discovering new materials was also demonstrated. Focusing on the map with $x\% = 75\%$ for the Co–Al–Ge ternary system, $Co_4Ge_{3.19}Al_{0.81}$ and $Co_2Al_{1.24}Ge_{1.76}$ were successfully synthesized.

## Acknowledgments
We thank Edanz (https://www.jp.edanz.com/ac) for editing an English draft of this manuscript. This work was supported by the Japan Science and Technology Agency (JST) CREST [Grant No. JPMJCR19J1] and the establishment of university fellowships towards the creation of science technology innovation from the JST [Grant No. JPMJFS2108].

## Data Availability
The Materials Project datasets were obtained from open database Materials Project https://next-gen.materialsproject.org/. The ICSD datasets were obtained from commercial database Inorganic Crystal Structure Database (ICSD) https://icsd.products.fiz-karlsruhe.de/. The data of predicted reactivity in this study are available via the website https://erm.starrydata.org/.